\documentclass[]{svmult}

\usepackage{makeidx}         
\usepackage{graphicx}        
\usepackage{multicol}        
\usepackage[bottom]{footmisc}

\makeindex             

\begin{document}
\newcommand{\beq}{\begin{equation}}
\newcommand{\eeq}{\end{equation}}
\newcommand{\mbf}{\mathbf}
\title*{Solar and Stellar Active Regions:Cosmic laboratories for the study of Complexity}
\author{Loukas Vlahos}
\institute{Department of Physics, University of Thessaloniki, 54124
  Thessaloniki, Greece
\texttt{vlahos@astro.auth.gr}}
%
%
\maketitle


\begin{abstract}
Solar active regions are driven dissipative dynamical systems. The turbulent convection zone  forces new magnetic flux tubes to rise above the photosphere and shuffles  the magnetic fields which are already above the photosphere. The driven 3D active region responds to the driver with the formation of Thin Current Sheets in all scales and releases impulsively  energy, when special thresholds are met, on the form of nano-, micro-, flares and large scale coronal mass ejections. It has been documented that active regions form self similar structures with area Probability Distribution Functions (PDF's)  following power laws and with fractal dimensions ranging from $1.2-1.7$. The energy release on the other hand follows a specific energy distribution law $f(E_T)\sim E_T^{-a}$, where $a \sim 1.6-1.8$ and $E_T$ is the total energy released. A possible explanation for the statistical properties of the magnetogrms and the energy release by the active region is that the magnetic field formation follows rules analogous to \textbf{percolating models}, and the 3D magnetic fields above the photosphere reach a \textbf{Self Organized Critical (SOC) state}. The implications of these findings on the acceleration of energetic particles during impulsive phenomena will briefly be outlined.
\end{abstract}

\section{Active Regions as driven non linear systems}
\label{sec:1}

The most energetic phenomena above the solar surface are associated with ``active regions (AR)". The 3D AR is a theater of intense activity of various (spatiotemporal) scales.  The 3D AR has a visual boundary at the photosphere (although its physical boundary, as we will see in the  the next section,  is inside the turbulent convection zone), and is subject to external forcing caused by the flux emergence from the solar interior and by the shuffling motions at the photosphere.

\begin{figure}[ht]
\centering
  \includegraphics[width=5cm]{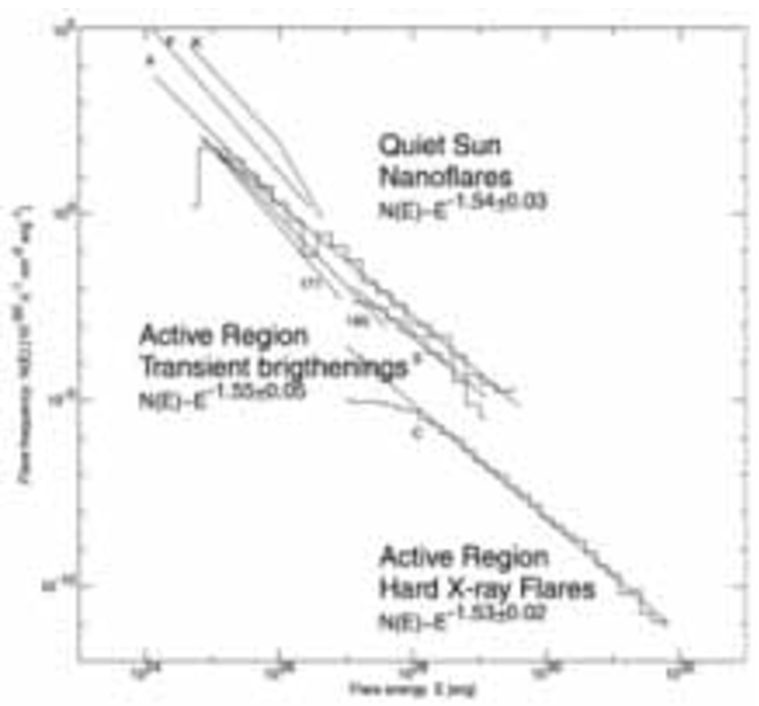}
  \includegraphics[width=5cm]{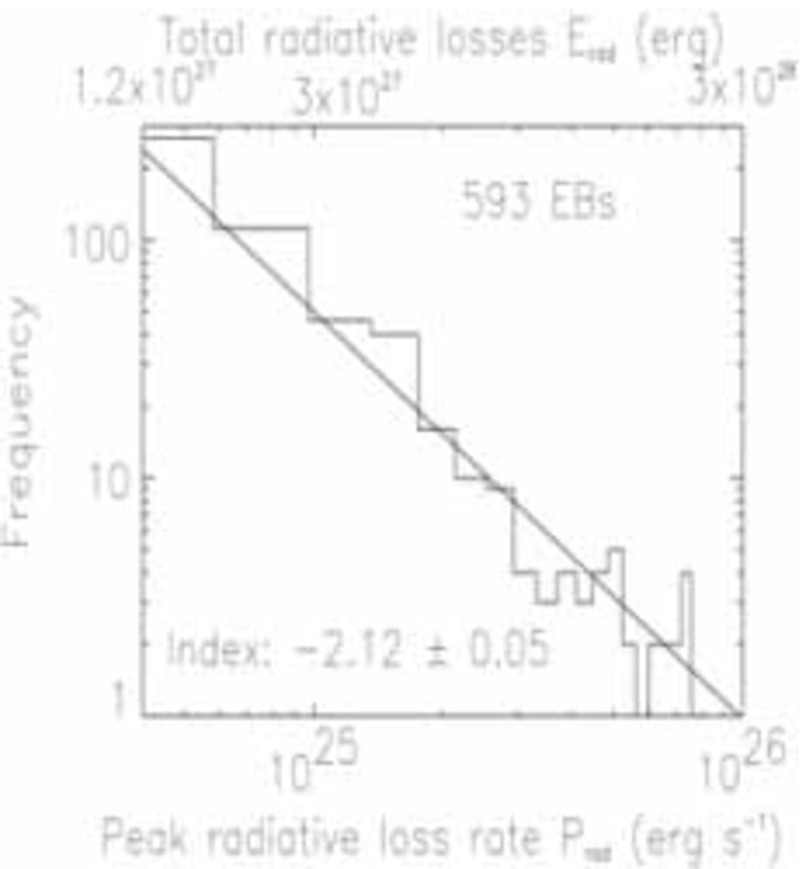}
  \includegraphics[width=6cm]{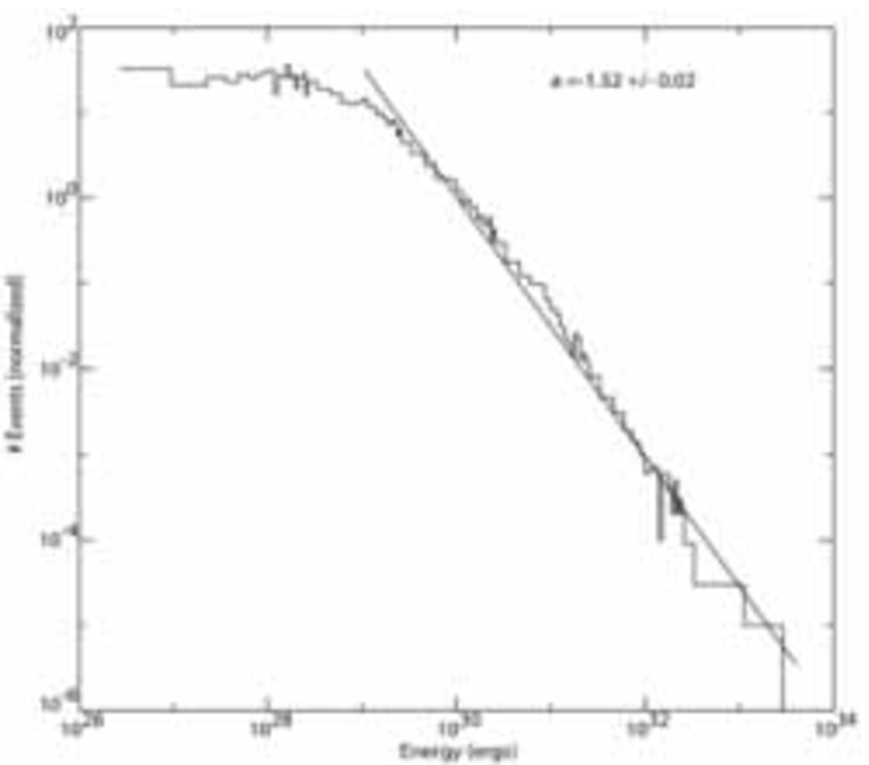}\\
  \caption{(a) Statistical properties of flares (see \cite{Markus}), (b) statistical properties of Ellerman Bombs (see \cite{geo02}), (c) statistical properties of the kinetic energy of CMEs (Courtesy of A. Vourlidas).}\label{fig:1}
\end{figure}

 Our main goal in this article is  to show that a variety of well known solar phenomena (e.g. coronal heating, flares, CME's, particle acceleration etc) can be understood in a unified manner by considering the solar active regions as externally (sub-photospherically) driven non linear systems. There are many well known statistical observations  which suggest that ARs are far from equilibrium: (1) The magnetic structures at the solar photosphere establish a \textbf{fractal} form and have power law size distributions, (2) the \textbf{explosive phenomena} (i.e nano-flares, micro-flares, flares and CME's) follow a very stable \textbf{power law frequency distribution} (3) the high energy particles, accelerated during solar flares, establish before leaving the accelerator, a \textbf{power law energy distribution} (see Fig. \ref{fig:1}).

 The main question addressed in this review then is: How an active region achieves all these statistical regularities? We argue  that the formation of AR's follows laws analogous to well known percolation models and simultaneously drives   the extrapolated 3D magnetic structures above the photosphere, leading the entire 3D structure to a Self Organized Critical (SOC) state.  \textbf{The coupling of two well known mechanisms of complex systems (percolation (as the driver) and self organized criticality for the energy dissipation) are behind all the observed regularities recorded on the data.}

\section{Active region formation: A percolating driver?}
\label{sec:2}
\begin{figure}[ht]
\centering
  \includegraphics[width=3.5cm]{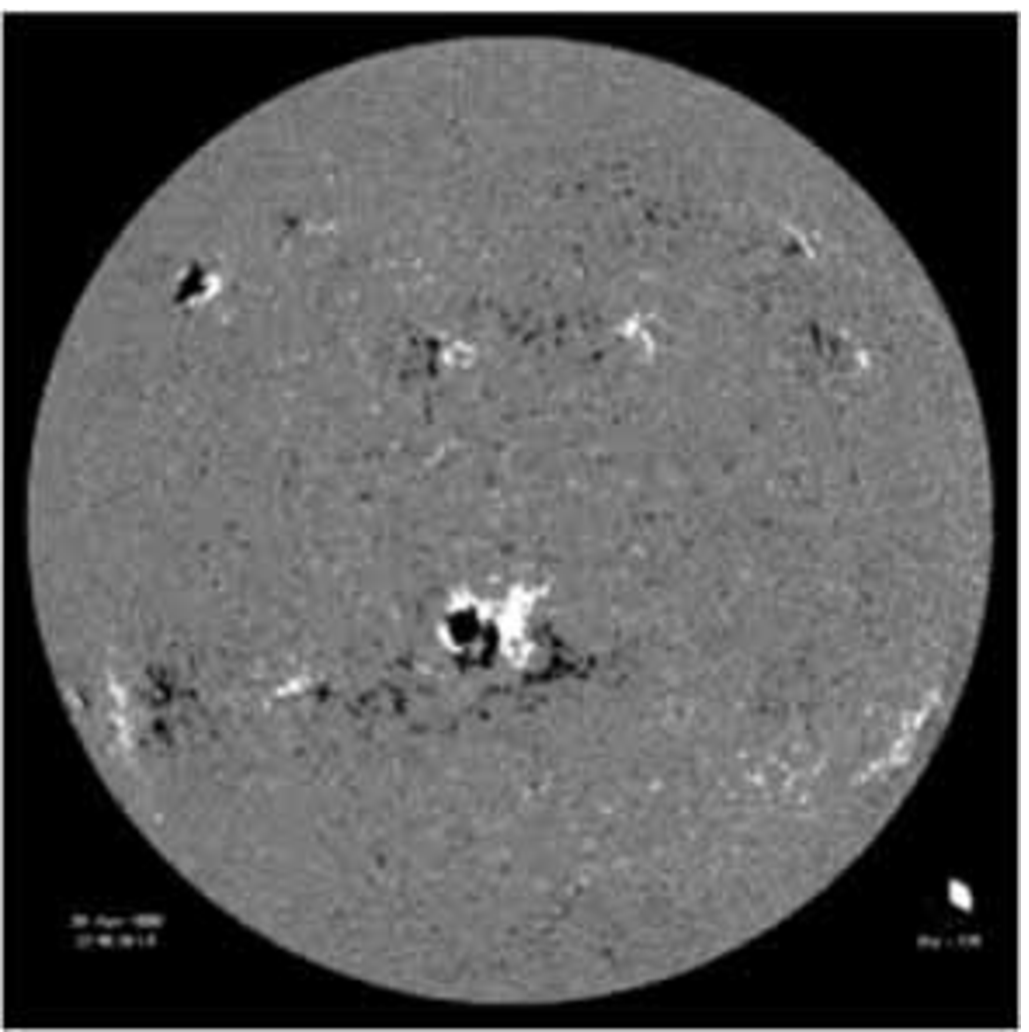}
  \includegraphics[width=4.5cm]{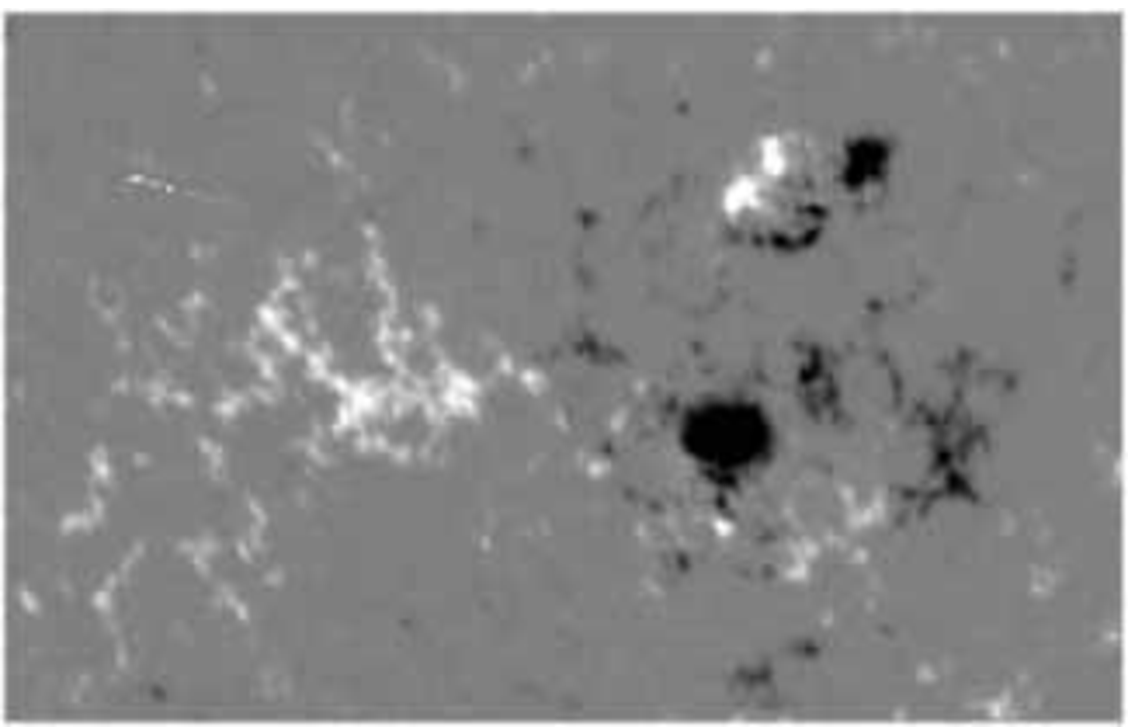}\\
  \caption{(a) Sunspots in a full disk magnetogram, (b) Magnetogram around an active region }\label{fig:2}
\end{figure}

We can learn a lot about the sub-photospheric activity by ``reading'' carefully the  magnetograms.  Both full disk and more detail magnetograms around specific AR are extremely useful tools (see Fig. \ref{fig:2}).   Two of their striking properties are found in the Probability Distribution Function (PDF) of their sizes  and their fractal properties.

Numerous observational studies have investigated the statistical
properties of active regions, using full-disc magnetograms.
These studies have examined among other parameters the size
distribution of  active regions, and their fractal dimension: {\em
The size distribution function} of the newly formed active regions
exhibits a well defined power law with index $\approx -1.94$
 and active regions cover only a small
fraction of the solar surface (around $\sim 8\%$) \cite{HarvZw93}.
  The \textit{fractal dimension} of the
active regions has been studied using high-resolution magnetograms
by \cite{Balke93}, and more recently by \cite{Meun99}. These
authors found, using not always the same method, a fractal
dimension $D_F$ in the range $1.2<D_F<1.7$.  Fractal dimensions in solar magnetic fields are typically calculated using the box-counting technique. The values of the fractal dimension depends on whether the structures themselves or just their boundaries are box counted. The analysis has been pursued even further using the concepts of multi-fractality. It is well known that an AR includes multiple types of structures such as different classes of sunspots, plages, emerging flux sub-regions, etc. The physics behind the formation and evolution of each of these structures is not believed to be the same, so the impact and the final outcome of convection zone turbulence in each of these configurations should not be the same. Numerous other tools have been used to uncover aspects of the complex behavior "mapped" by the convection zone on the photospheric boundary,  e.g. generalized correlation dimension, structure formations, wavelet power spectrum \cite{geo05}.

Theoretical studies on the formation of AR can be divided in two main categories: (1)  The evolution of one or two slender and isolated flux tubes (see recent articles on this topic \cite{Amari, kliem, aulanie2,Archontis,Gal3}), (2) using standard percolation techniques. We stress the second method here since  we place special emphasis on the complexity of AR.

A percolation model was
proposed to simulate the formation and evolution of active regions at the photosphere in
\cite{Wen92,Sei96}. In this model, the evolution of the magnetograms is followed by reducing all the complicated convection zone dynamics into three dimensionless parameters.
The emergence and evolution of magnetic flux on the solar surface
in the  2-D cellular automaton (CA) which  is probabilistic and is based on the competition between
two ``fighting" tendencies: \textbf{stimulated} or
\textbf{spontaneous} emergence of new magnetic flux, and the
disappearance of flux due to \textbf{diffusion} (i.e.\ dilution
below observable limits), together with random \textbf{motion} of the
flux tubes on the solar surface (this processes mimics the shuffling). This percolation
model explains the observed size distribution of active regions
and their fractal characteristics \cite{Meun99, vla02a}. It was later used for the reconstruction of 3-D active regions using the force free approximation and many of the observational details reported in \cite{vla04} were reproduced \cite{Fragos03}.

\section{Are AR in a self organized critical state?}\label{sec:3}
\subsection{3-D extrapolation of magnetic field lines
and the formation of Unstable Current Sheets} \label{subsec:extra}
The energy needed to power solar flares is provided by
photospheric and sub-photospheric motions and is stored in
non-potential coronal magnetic fields. Since the magnetic
Reynolds number is very large in the
solar corona, MHD theory states that \textbf{magnetic energy
can only be released in localized regions where the magnetic field forms
small scale structures and steep gradients, i.e. in thin current sheets (TCS)}.

Numerous articles (see recent reviews \cite{Dem, Longc}) are devoted
to the analysis of magnetic topologies which can host TCSs.
The main trend of current research in this area is to find ways to
realistically reconstruct the 3-D magnetic field topology in the corona
based on the available magnetograms and large-scale plasma motions
at the photosphere.
A realistic magnetic field generates many ``poles and sources"
\cite{Longc} and naturally has a relatively large number of TCSs. We
feel that detailed representations inside the  3D AR of the
TCS's are mathematically appealing only for relatively simple magnetic
topologies (dipoles, quadrupoles, symmetric magnetic arcades \cite{aulanie1}).
When such simple topologies are broken in the photosphere, for example due to large-scale sub-Alfv\'enic
photospheric motions or the emergence of new magnetic flux that disturbs the
corona, such tools may be less useful.
All these constraints restrict our ability to reconstruct fully the
dynamically evolving magnetic field of an active region (and it is not clear
that an exact reconstruction will ever be possible).

Many of the widely used magnetograms measure only the line of sight
component of the magnetic field. The component of the
magnetic field vertical to the surface matches the measured
magnetic field only at the center of the disk and becomes
increasingly questionable as the limb is approached.
Extrapolating the measured magnetic field is
relatively simple if we assume that the magnetic field is
in force-free equilibrium:
\begin{equation}\label{ff}
    \nabla \times \vec{B}=\alpha(\vec{x}) \vec{B}
\end{equation}
where the function $\alpha(\vec{x})$ is arbitrary except for the
requirement $\textbf{B} \cdot \nabla \alpha(\vec{x})=0$, in order to preserve
$\nabla \cdot \vec{B}=0$.
Eq. (\ref{ff}) is non-linear since both $\alpha(\vec{x})$ and
$\vec{B}(\vec{x})$ are unknown. We can simplify the
analysis of Eq. \ref{ff} when  $\alpha$=constant. The solution is
 easier still when $\alpha=0$, which is equivalent to
assuming the coronal fields to contain no currents (potential field),
hence no free energy, and thus they are uninteresting.
\begin{figure}[ht]
\centering
  \includegraphics[width=8cm]{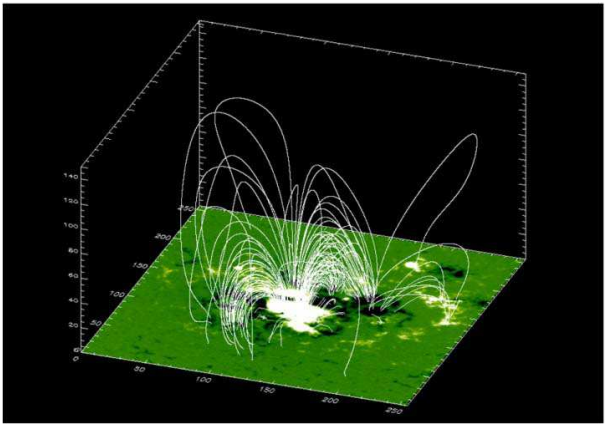} \\
  \caption{Force free extrapolation of magnetograms and reconstruction of the magnetic fields inside an AR}\label{fig:3}
\end{figure}
A variety of techniques have been developed for the reconstruction
of the magnetic field lines above the photosphere and the search for
TCSs \cite{Longc}. It is beyond the scope of this article to
discuss these techniques in detail. For instructive purposes,
we use the simplest method available, a linear
force free extrapolation, and search for ``sharp'' magnetic
discontinuities in the extrapolated magnetic fields as in
Vlahos and Georgoulis \cite{vla04}: They use an observed active-region
vector magnetogram and then: (i) resolve the intrinsic azimuthal
ambiguity of $180^o$ \cite{geo04a}, and
(ii) find the best-fit value $\alpha _{AR}$ of the
force-free parameter for the entire active region, by minimizing the difference
between the extrapolated and the ambiguity-resolved observed
horizontal field (the ``minimum residual'' method of \cite{Leka03}).
They perform a linear force-free extrapolation
\cite{al81} to determine the three-dimensional magnetic
field in the active region (see Fig. \ref{fig:3}). Although it is known that magnetic fields at
the photosphere are not force-free \cite{geo04b},
they argue that a linear force-free approximation is suitable for
the statistical purposes of their study.

Two different selection criteria were used in order to identify
potentially unstable locations (identified as
the afore-mentioned TCSs) \cite{vla04}.
These are (i) the Parker angle, and (ii) the total magnetic field gradient.
The angular difference $\Delta \psi$ between two adjacent magnetic field
vectors, $\mbf{B_1}$ and $\mbf{B_2}$, is given by $\Delta \psi =
\cos^{-1} [\mbf{B_1} \cdot \mbf{B_2}/(B_1 B_2)]$. Assuming a cubic
grid, they estimated six different angles at any given location, one
for each closest neighbors. The location is
considered potentially unstable if at least one $\Delta \psi _i >
\Delta \psi _c$, where $i \equiv \{ 1,6 \}$ and $\Delta \psi _c =
14^o$. The critical value $\Delta \psi _c$ is the Parker
angle which, if exceeded locally, favors tangential discontinuity formation
and the triggering of fast reconnection \cite{Par83,Parker88}. In
addition, the total magnetic field gradient between two adjacent
locations with magnetic field strengths $B_1$ and $B_2$ is given by
$|B_1 - B_2| / B_1$. Six different gradients were calculated at any
given location. If at least one $G_i > G_c$, where $i \equiv \{ 1,6
\}$ and $G_c=0.2$ (an arbitrary choice), then the location is
considered potentially unstable. When a TCS obeys
one of the criteria listed above, it will be transformed to an Unstable Current
Sheet (UCS).
Potentially unstable volumes are formed by the merging of adjacent
selected locations of dissipation. These volumes are given by
$V=N\lambda^2 \delta h$, where
$N$ is the number of adjacent locations, $\lambda$ is the pixel size
of the magnetogram and $\delta h$ is the height step of the
force-free extrapolation. The free magnetic energy $E$ in any volume
$V$ is given by

\beq E= {{\lambda ^2 \delta h} \over {2 \mu_0}} \sum
_{l=1} ^N (\mbf{B_{ff}}_l- \mbf{B_p}_l)^2 \label{energy} \eeq
where
$\mbf{B_{ff}}_l$ and $\mbf{B_p}_l$ are the linear force-free
and the potential fields at location $l$ respectively. The
assumption used is that any deviation from a potential configuration
implies a non-zero free magnetic energy which is likely to be
released if certain conditions are met.
\begin{figure}
\centering
\includegraphics[height=5cm]{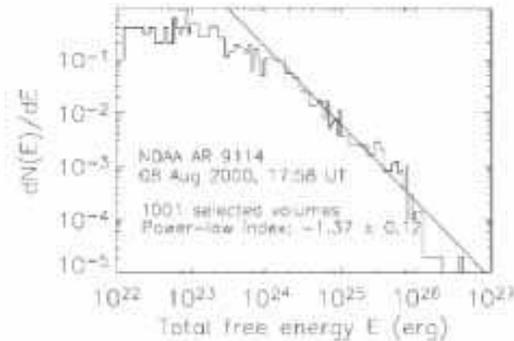}
%
\caption{Typical distribution function of the total free energy in
the selected volume, on using a critical  angle $14^\circ$
\cite{vla04}.\label{Vlahos1}}

\end{figure}
UCS are created naturally in active regions even during their
formation  and the free energy available
in these unstable volumes follows a \textbf{power law distribution}
with a well defined exponent (Fig. \ref{Vlahos1}).
Vlahos and Georgoulis concluded that active regions store energy in many unstable locations,
forming UCS of all sizes (i.e. the UCS have a self-similar structure).
The UCS are fragmented and distributed
inside the global 3-D structure. Viewing the flare in the context of
the UCS scenario presented above, we can expect, depending of the size distribution and
the scales of the UCS, to have flares of all sizes.

\subsection{A Cellular Automata model for the energy release in AR} \label{subsec:CA}
Coronal energy release observed at various wavelengths shows impulsive behaviour with
events from flares to bright points exhibiting intermittency in time and space. Intense X-ray flare emission typically lasts several minutes to tens of minutes, and only a few flares are recorded in an active region that typically lives several days to several weeks. The flaring volume is small compared to the volume of an active region, regardless of the flare size. Intermittency is the dynamical response of a turbulent system when the triggering of the system is the result of a critical threshold for the instability \cite{Cassak}. In a turbulent system one also expects \textbf{self organization}, i.e. the reduction of the numerous physical parameters (degrees of freedom) present in the system to a small number of significant degrees of freedom that regulates the system's response to external forcing \cite{nicolis}. This is the reason for the success of concepts such as Self Organized Criticality (SOC) \cite{Bak87, Bak96} in explaining the statistical behavior of flares.   Cellular Automata (CA) models typically employ one variable (the magnetic field, vector potential, etc) and study its evolution subject to external perturbations. \textbf{When a critical threshold is exceeded (when the TCS becomes an UCS), parts of the configuration are unstable, and will be restructured to re-establish stability.} The rearrangement may cause instabilities in adjacent locations, so the relaxation of the system may proceed as an avalanche-type process. In SOC flare models \cite{Lu91,Lu93, Vla95} each elementary relaxation is viewed as a single magnetic reconnection event, so magnetic reconnection is explicitly assumed to occur with respect to a critical threshold.

In solar MHD an UCS disrupts either when its width becomes smaller than a critical value \cite{Pets}, or when the magnetic field vector forms tangential discontinuities exceeding a certain angle \cite{Parker72}, or when magnetic field gradients are steep enough to trigger restructuring \cite{pr03}. We noticed that a critical threshold is involved in all cases: the first process points to the turbulent evolution in the magnetic field configuration and the onset of anomalous resistivity, while the latter two imply magnetic discontinuities caused either by the orientation of the magnetic field vector or by changes of the magnetic field strength. Magnetic field gradients and discontinuities imply electric currents via Amp\'ere's law, so a critical magnetic shear or gradient implies a critical electric current accumulated in the current sheet which in turn leads to the onset of anomalous resistivity \cite{Papadop77,Par83}.

\begin{figure}[ht]
  \centering
\includegraphics[width=10cm]{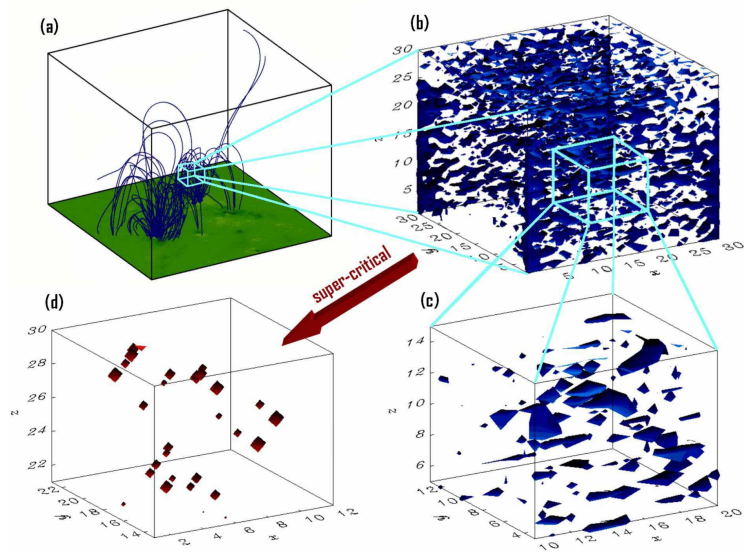}
  \caption{(a) Simulated magnetogram of a photospheric active region
   and force-free magnetic field-lines, extrapolated into the corona
       (b) Sub-critical current iso-surfaces in space, as yielded by the X-CA
   model, which models a sub-volume of a coronal active region. ---
   (c) The same as (b), but zoomed.
   (d) Temporal snap-shot of the X-CA model during a flare, showing the
   spatial distribution of the UCS (super-critical current iso-surface)
   inside the complex active region \cite{vla05}.}\label{plot1}
\end{figure}

One way of modeling the appearance, disappearance, and spatial organization
of UCS inside a large-scale topology is with the use of the Extended Cellular
Automaton (X-CA) model \cite{Isl98,Isl00,Isl01}. Fig. \ref{plot1}
illustrates some basic features of the X-CA model.
The X-CA model has as its core a cellular automaton model of the
sand-pile type and is run in the state of Self-Organized Criticality
(SOC). It is extended to be fully consistent with MHD: the primary
grid variable is the vector-potential, and the magnetic field and
the current are calculated by means of interpolation as derivatives
of the vector potential in the usual sense of MHD, guaranteeing
$\nabla \cdot \vec B=0$ and $\vec J=(1/\mu_0)\nabla\times\vec B$
everywhere in the simulated 3-D volume. The electric field
is defined as $\vec E=\eta \vec J$, with $\eta$ the diffusivity. The
latter usually is negligibly small, but if a threshold in the
current is locally reached ($\vert\vec J\vert>J_{cr}$), then current-driven
instabilities are assumed to occur, $\eta$ becomes anomalous
in turn, and the resistive electric field locally increases
drastically. These localized regions of intense electric fields are
the UCS in the X-CA model.

The X-CA model yields distributions of total energy and peak flux
which are compatible with the observations. The UCSs in the X-CA form
a set which is highly fragmented in space and time: the individual
UCS are small scale regions, varying in size, and are short-lived.
They do not form in their ensemble a simple large-scale structure,
but form a fractal set with fractal dimension roughly
$D_F=1.8$ \cite{vla05}. The individual UCS also do not usually split
into smaller UCS, but they trigger new UCSs in their neighborhood, so that
different chains of UCS travel through the active region, triggering
new side-chains of UCS on their way. It is obvious that the rules of this simulation do not include the fragmentation of the UCS, in many ways through the results coincide with the MHD simulations \cite{Gal2}.

\section{Active Regions  as multi-scale physics laboratories}

So far we have discussed very briefly: (1) the formation of an AR as it is mapped in the magnetogram, (2) the use of the magnetogram as non-linearly evolving driver for the 3D AR, (3) the reconstruction of the 3D AR using simple techniques and the search for Thin Current Sheets (TCS)  where energy may be dissipated. The critical transition of a TCS to a rapidly reconnecting structure (UCS) is essential  for the 3D AR to reach a SOC state. The TCS formed inside the AR extend from the large scales ($10^{10}$ cm) which are very unstable and rapidly fragment down to a few meters (on the order of the ion gyro radius)  where the fast reconnection ignites. We have already discovered on all these levels enormous complexity. The main question now is: Do the UCS remain stable and dissipate magnetic energy or are they fragmented even further?

Onofri et al. \cite{marco} studied the nonlinear evolution of
current sheets using the 3-D incompressible and dissipative MHD equations in a slab
geometry. The nonlinear evolution of the system is characterized by the
formation of small scale structures, especially in the lateral
regions of the computational domain, and coalescence of magnetic
islands in the center.

\begin{figure}
\centering
  \includegraphics[width=5cm]{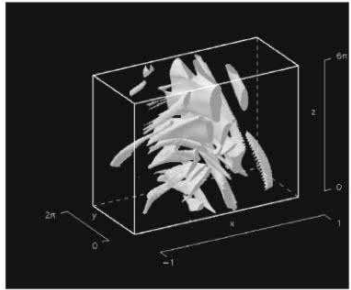}
  \caption{Current isosurfaces showing the formation of current filaments, \cite{marco}}\label{onofri}
\end{figure}

This behavior is reflected in the 3-D structure of the current (see Fig. \ref{onofri}), which shows that the initial equilibrium is destroyed by the formation of current filaments, with a prevalence of small scale features.  The final stage of these simulations is a turbulent state, characterised by many spatial scales, with small structures produced by a cascade with wavelengths decreasing with increasing distance from the current sheet. In contrast, inverse energy transfer leads to the coalescence of magnetic islands producing the growth of two-dimensional modes. The energy spectrum approximates a power law with slope close to 2 at the end of the simulation \cite{marco1}.
Similar results have been reported by many authors using several approximations \cite{Drake06, Milano, Larosa, Shibata01}. It is also interesting to note that similar results are reported from magnetic fluctuations in the Earth's magnetotail \cite{Zelenyi}.

We have now created current structures on all scales and the next question is how particles will respond to the electric fields developed at all these scales due to the presence of enhanced resistivity at the small scales? The problem of particle acceleration is beyond the domain of MHD theory or even the two-fluid description of the plasmas. Only kinetic theory can follow the evolution of the particle distribution inside a fractally distributed electric field environment.  Particle (anomalous) diffusion and the build up of non-thermal velocity distributions in localized structures distributed inside the 3D AR was the subject of many recent articles \cite{vla05} (see also a recent review \cite{vla08}).

\section{Conclusions and Summary}
In this review we have attempted to show that many well known solar phenomena, treated separately in many recent reviews (e.g. coronal heating, flares, CME's and particle acceleration) are symptoms of the formation and evolution of  ARs.
We can now pose the question: How an AR works? We propose here that four main steps are crucial.
\begin{itemize}
  \item \textbf{Step 1: The driver: } The photospheric activity with emerging magnetic flux and photospheric flows are the main ingredients of the driver. The main rules of the evolution of an active region can be explained by using a simple \textbf{percolation model}.
  \item \textbf{Step 2: The 3D magnetic "skeleton"  and the storage of magnetic energy:} This is a very difficult task and remains open challenge. Simple forms of extrapolation show that the storage of magnetic energy is in Thin Current Sheets (TCS) which are formed at all scales and follow very interesting statistical regularities (\textbf{Self organization of the storage of magnetic energy}).
  \item \textbf{Step 3: Eruptions and Criticality:} The energy release is possible only when the TCS become very thin and move to the state of Unstable Current Sheets (UCS). From  small scale  and \textbf{confined eruptions} up to very large scales reorganization of the corona that is followed by \textbf{flares and CMEs,} the eruptions are viewed here as ``avalanches" coming out of a system which is in \textbf{Self Organized Critical state (SOC)}.
  \item \textbf{Step 4: Particle Acceleration:} The \textbf{fractal distribution in 3D-space} of the energy release sites followed naturally from a space filling \textbf{fractal distribution of E-fields}. Particles \textbf{"Diffuse" inside a  Network }of accelerators and the  "\textbf{Accelerator}"  is distributed over a relatively large volume, as is the distribution of the energy release sites. How particles are accelerated in \textbf{Networks of E-fields} is an interesting statistical mechanics problem.
\end{itemize}

In Summary we conclude that the complexity of the magnetograms, the formation of millions of TCS which, after passing a  threshold become UCS at sporadic places inside the AR, and the further fragmentation of all UCS practically to all scales, play an important role in the formation of high energy particles. Therefore in 3D ARs of all scales are active from the very large scales (thousands of Kilometers, treated by MHD) down to meters (treated by kinetic equations). This multi-scale and complex environment still maintains many interesting statistical regularities which are manifested   in the numerous statistical laws recorded from the data so far.

\begin{acknowledgement}
 I would like to thank my colleagues Drs H. Isliker, M. Georgoulis and Mr T. Fragos  for many useful conversations.
\end{acknowledgement}
\end{document}